\documentclass{pasj00}

\newcommand{\Lji}{L_{\rm j}^{\rm iso}}
\newcommand{\Lj}{L_{\rm j}}
\newcommand{\Ljie}{L_{\rm j,52}^{\rm iso}}
\newcommand{\Lje}{L_{\rm j,51}}

\begin{document}
\title{Nuclear Composition of Magnetized GRB Jets}
\author{Sanshiro \textsc{Shibata}\altaffilmark{1} and Nozomu \textsc{Tominaga}\altaffilmark{1,2}}  
\altaffiltext{1}{Department of physics, Konan University, 8-9-1 Okamoto, Kobe, Hyogo
  658-8501, Japan}
\altaffiltext{2}{Kavli Institute for the Physics and Mathematics of the Universe,
  University of Tokyo, 5-1-5 Kashiwanoha, Kashiwa, Chiba 277-8583, Japan}
 \email{E-mail: d1221001@center.konan-u.ac.jp}


\maketitle

\begin{abstract}
  We investigate
  the fraction of metal nuclei in the relativistic jets of gamma-ray bursts
  associated with core-collapse supernovae. We simulate the fallback in 
  jet-induced explosions with two-dimensional relativistic hydrodynamics 
  calculations and the jet acceleration with steady, radial, relativistic 
  magnetohydrodynamics calculations, and derive detail nuclear 
    composition of the jet by postprocessing calculation.
  We found that if the temperature at the jet launch site is above 
  $4.7\times 10^9$K, quasi-statistical equilibrium (QSE) is established and 
  heavy nuclei are dissociated to light particles such as $^4$He during the
  acceleration of the jets. The criterion for the survival of metal nuclei is
  written in terms of the isotropic jet luminosity as 
  $L_{\rm j}^{\rm iso} \lesssim 3.9\times 10^{50}(R_{\rm i}/10^7{\rm cm})^2
  (1+\sigma_{\rm i})~{\rm erg~s^{-1}}$, where $R_{\rm i}$ and $\sigma_{\rm i}$
  are the initial radius of the jets and the initial magnetization parameter,
  respectively. If the jet is initially dominated by radiation field (i.e., 
  $\sigma_{\rm i} \ll 1$) and the isotropic luminosity is relatively high
  ($L_{\rm j}^{\rm iso}\gtrsim 4\times 10^{52}~{\rm erg s^{-1}}$), the metal
  nuclei cannot survive in the jet.
  On the other hand, if the jet is mainly accelerated by magnetic field (i.e.,
  $\sigma_{\rm i} \gg1$), metal nuclei initially contained in the jet can
  survive without serious dissociation even for the case of high luminosity
  jet. If the jet contains metal nuclei, the dominant nuclei are $^{28}$Si, 
  $^{16}$O, and $^{32}$S and the mean mass number can be 
  $\langle$A$\rangle\sim25$.

\end{abstract}

\section{INTRODUCTION}
Gamma-ray bursts (GRBs) are one of the most energetic phenomena in the
universe. They radiate enormous energies of the order of $10^{51}{\rm ~ergs}$
mainly in the form of gamma-rays with the short duration, typically of
$\sim10{\rm ~s}$. Although the detailed radiation mechanism of gamma-rays is 
under debate, the radiation is thought to be originated from the
ultra-relativistic jets with $\Gamma\gtrsim 100$, where $\Gamma$ is the
Lorentz factor of the jets (e.g., \cite{Meszaros06}).  Furthermore, it is
known that GRBs with duration longer than $2{\rm ~s}$, called long GRBs,
are associated with highly energetic type Ic supernovae (e.g.,
\cite{Iwamoto98}). This implies that the relativistic jets are launched at deep
inside of massive progenitor stars after the onset of the core-collapse.

Nuclear composition of GRB jet have been studied by several authors.
\citet{Pruet02}, \citet{Lemoine02} and \citet{Beloborodov03} investigated
within the framework of the standard fireball model (e.g., \cite{Meszaros06}), 
assuming that GRB jet initially consists of free nucleons. The nucleons can
recombine into deuterium and/or $\alpha$ particles as the jet expands and
cools but metal nuclei heavier than carbon can not be produced. Thus, they
conclude that the GRB jet consists only of light nuclei.

However, there is a caveat that the initial composition of the jet is not
necessarily dominated by the free nucleons because  the mechanism to
launch a well-collimated relativistic jet is not specified. The proposed
mechanisms include neutrino annihilation (e.g., \cite{Woosley93,MW99}) and
magnetic field (e.g., \cite{BZ77,Brown00,McKinney06}). In fact, some
observations of GRBs suggest that their relativistic jets were initially
dominated by the magnetic field energy flux (e.g.,
\cite{ZP09,Guiriec11}). When the relativistic jets are launched by the
magnetic field, the jets could involve heavy nuclei such as $^{56}{\rm Ni}$
\citep{Fujimoto08} and thus the nuclear composition of the jets may be
different from the one that is derived within the framework of standard
fireball model.

 \citet{Horiuchi12} investigated survival of metal nuclei in relativistic
jets. They analytically estimated the  conditions to survive
photodisintegration and spallation at the base of the jets, in the
  accelerating jets, and at the emission region of GRB. They argued that the
  nuclei can avoid the destructions for a range of jet parameters. 
 For examples, the metal nuclei can survive at the base of the jet
if the radius of the central engine is greater than $\sim10^8~{\rm cm}$ or if
the radiation luminosity is less than $\sim10^{48}~{\rm erg/s}$.
They also investigated the possibility for
entrainment of metal nuclei from stellar material during the jet propagation
in the progenitor star and suggested that the entrainment is possible 
depending on the model parameters.
However, they did not calculate detail nuclear reactions and did not
present final nuclear compositions.

In this paper, we calculate non-equilibrium nuclear reactions during the
fallback and jet acceleration, based on the thermal histories derived by the
relativistic hydrodynamics calculations and the steady, radial, relativistic
magnetohydrodynamics calculation. Then we present the criterion for the metal 
nuclei to survive without serious dissociation.
For the initial composition, we assume that the GRB jet is initially made from
falling matter during a relativistic jet-induced explosion.
We adopt Wolf-Rayet
stars proposed to be progenitors of GRB-SN and circumstantially treat the
jet acceleration by thermal and/or magnetic pressure gradient and
nucleosynthesis in relativistic jets.

This paper is organized as follows. In Section \ref{sec:method}, we describe
the method and the model of GRB jet. In Section \ref{sec:results}, we present
fallback in jet-induced explosions, hydrodynamic properties of the 
accelerating jet, and the final nuclear composition of the 
jet.  The conclusion is presented in Section
  \ref{sec:conclusion}. Finally, the discussions are presented in Section
  \ref{sec:discussion}.

\section{METHOD \& MODEL}\label{sec:method}
\subsection{Method}

We calculate the nuclear composition of magnetized GRB jets with
following two steps: (1) We follow fallback in a relativistic jet-induced
explosion with a two-dimensional relativistic hydrodynamics calculation
in order to derive the initial composition of jets. (2) We follow
the acceleration of magnetized GRB jets after the launch of jets with a
steady, radial, relativistic magnetohydrodynamics
calculation, assuming an interaction between the jet and stellar mantle
does not influence on the acceleration of the GRB outflow. 

According to thermodynamical histories during the fallback and
acceleration, nuclear reactions are calculated as a postprocessing
(e.g., \cite{hix96,hix99}). The nuclear reaction network includes 281
isotopes up to $^{79}{\rm Br}$. Here, we adopt an equation of state for
relativistic particles, $p=e_{\rm int}/3$, which means $\gamma_{\rm a}=4/3$ 
where $\gamma_{\rm a}$ is the adiabatic index, and assume that the internal
energy is dominated by contribution from photons and $e^{\pm}$
pairs. Temperature $T$ is derived by following equation
(e.g., \cite{Freiburghaus99}):
\begin{equation}\label{eq:e_int}
 e_{\rm int}=aT^4 \left( 1+\frac{7}{4}\frac{T_9^2}{T_9^2+5.3} \right),
\end{equation}
where $a=7.57\times 10^{-15}~{\rm erg~cm^{-3}~K^{-4}}$ is the radiation
constant and $T_9=T/(10^9 ~{\rm K})$. We note that the non-relativistic gas
pressure is negligible compared with radiation pressure because of the high
entropy.  

\subsubsection{Fallback in relativistic jet-induced explosions}

In jet-induced explosions, a considerable fallback takes place along the
equatorial plane (e.g., \cite{mae03b}) and thus an interaction between
the jets and cocoon and the stellar mantle determines which mass elements
fallback to a central remnant. Therefore, a numerical simulation is required
to correctly treat fallback in explosions with relativistic jets. Hence, we 
calculate relativistic jet-induced explosions of C+O stars with the use
of a two-dimensional relativistic Eulerian hydrodynamic code with the
Newtonian self-gravity \citep{Tominaga07,Tominaga09}. 

Since the GRB jets consist of the falling matter, the initial
composition of the outflow is set to be an integration of falling
matter. However, as the freefall
time of the matter at the inner boundary is short ($<0.1$~s), 
materials, that fall through the inner boundary well before the
initiation of the jet injection, are likely to have been accreted to the
central remnant before the jet injection and not to be re-ejected
as the relativistic jets. Therefore,
in this paper, only the matter falling after the initiation of the jet
injection are assumed to be re-ejected and integrated as the initial
composition of the outflow. 

As the material falls, the temperature increases and nucleosynthesis may
take place. Nucleosynthesis during the infall from
the presupernova location to $r=R_{\rm i}$ is calculated with a
thermodynamical history taking into account
heating due to the infall and cooling due to the Urca process
\citep{B02}. 

\subsubsection{Steady relativistic magnetized outflow}\label{sec:formulation}

We treat a GRB jet as a steady, radial, magnetized outflow with
efficient magnetic dissipation. We assume that magnetic fields in the outflow
are dominated by a toroidal component and that the field efficiently
dissipates via magnetic reconnection as the outflow expands. Such efficient
dissipation of magnetic fields creates strong magnetic pressure gradient which
enables a direct conversion of magnetic field energy into kinetic
energy \citep{Drenkhahn02,DS02}. We neglect gravitational force because the 
gravitational energy is small compared with the radiation energy or magnetic
field energy in the models considered here.

The mass, momentum, and energy conservation equations for the steady,
radial outflow with a toroidal magnetic field are described as follows
(e.g. \cite{LB03}):
\begin{equation}\label{eq:mass}
\frac{1}{r^2} \frac{\partial }{\partial r} \left(r^2 \rho \Gamma v\right)=0,
\end{equation}
\begin{equation}\label{eq:mome}
\frac{1}{r^2} \frac{\partial }{\partial
  r}\left[r^2\left\{\left(w+b^2\right)\Gamma^2
v^2+\frac{b^2}{2}\right\}\right]+\frac {\partial p}{\partial r}=0,
\end{equation}
\begin{equation}\label{eq:ener}
\frac{1}{r^2} \frac{\partial }{\partial r}\left(r^2 (w+b^2) \Gamma^2
v\right)=0,  
\end{equation}
where $r$, $\rho$, $\Gamma$, $v$, $p$, $w$, and $b$ are
distance from the center, proper mass density, outflow Lorentz factor, outflow
velocity, pressure, enthalpy, and toroidal component of the magnetic
four-vector, respectively. The enthalpy is defined by $w=\rho+e_{\rm int}+p$
where $e_{\rm int}$ is the internal energy. Here we adopt the unit system in
which the speed of light is unity. 

We employ a following evolution equation for the magnetic field which includes
dissipation of non-axisymmetric magnetic field produced by inclined
rotator \citep{Drenkhahn02}:
\begin{equation}\label{eq:magn}
\frac{\partial }{\partial r}\left(r b \Gamma v \right)=-\frac{r\Gamma
  b}{\tau_{\rm dis}},
\end{equation}
where $\tau_{\rm dis}$ is the dissipation time scale derived as follows:
\begin{equation}\label{eq:timescale}
\tau_{\rm dis} = \frac{2\pi \Gamma^2}{\epsilon \Omega}{\sqrt{1+u_{\rm A}^{-2}}},
\end{equation}
where $u_{\rm A} = b/\sqrt{w}$ is Alfv$\acute{\rm e}$n four-velocity, $\Omega$
is an angular frequency of central object, and $\epsilon$ is a dimensionless
factor.

 From Equation (\ref{eq:mass})-(\ref{eq:timescale}), the evolution of the Lorentz
factor can be written as
\begin{equation}\label{eq:acce}
\frac{\partial \Gamma}{\partial r}=\frac{\gamma_{\rm
    a}v^2\Gamma^3}{(\Gamma^2-\Gamma^2_{\rm f})(w-\gamma_{\rm
    a}p)}\left\{\frac{2p}{r} +\frac{(2-\gamma_{\rm a})b^2}{\gamma_{\rm
    a}v\tau_{\rm dis}}
\right\},
\end{equation}
where $\Gamma_{\rm f}$ is a Lorentz factor corresponding to phase velocity of 
a fast
magnetosonic wave and can be expressed as (e.g., \cite{LB03})
\begin{equation}
\Gamma_{\rm f}=\sqrt{\frac{\gamma_{\rm a}p+b^2}{w+b^2}}.
\end{equation}
 Equation (\ref{eq:acce}) indicates that the
flow is accelerated only when $\Gamma>\Gamma_{\rm f}$ and the acceleration of
the flow is infinity when $\Gamma=\Gamma_{\rm f}$. These behaviors of
the flow is due to the absence of the gravity in the above
formulation. Thus, although the magnetic field possesses a part of the
total energy, we assume that the flow is accelerated as if
there is no magnetic field and the magnetization parameter defined by
$\sigma\equiv b^2/w$ is constant until $\Gamma$ reaches $2\Gamma_{\rm f}$.
Then, the flow is accelerated with Equation (\ref{eq:acce})  at
 $\Gamma<\Gamma_{\rm f}$.

\subsection{Model}

We parameterize the GRB outflow with six parameters: isotropic energy
deposition rate $\Lji$, initial radius of the outflow
$R_{\rm i}$, initial Lorentz factor $\Gamma_{\rm i}$, maximum Lorentz factor
$\Gamma_{\rm max}$, angular frequency of central object (including
dimensionless factor) $\epsilon \Omega$, and initial magnetization parameter
$\sigma_{\rm i} = b_{\rm i}^2/w_{\rm i}$,
where $b_{\rm i}$ and $w_{\rm i}$ is the initial  troidal component of
  the magnetic four-vector and the
initial enthalpy, respectively. 

\begin{table*}
\caption{Summary of the model parameters}
\begin{center}
\begin{tabular}{ccccccc}
  \hline
  Model & $L_{\rm j}^{\rm iso}$ & $L_{\rm j}$ & Z & $\Gamma_{\rm max}$ &
  $\Gamma_{\rm i}$ & $\epsilon \Omega$\\
   & ($10^{52}~{\rm erg~s^{-1}}$)& ($10^{51}~{\rm erg~s^{-1}})$ & & & & (${\rm s^{-1}}$)\\
  \hline
  \hline
  A & 4 & 1.5 & 0 & 100 & 1.23 & $10^{3}$\\
  \hline
  B & 10 & 3.4 & 0 & 100 & 1.23 &$10^{3}$\\
  \hline
  C & 4 & 1.5 &0.02 & 100 & 1.23 &$10^{3}$\\
  \hline
  D & 10 & 3.4 & 0.02 & 100 & 1.23 &$10^{3}$\\
  \hline
\end{tabular}
\label{tab:model}
\end{center}
\end{table*}

In this paper, we investigate the dependence of final nuclear
composition on $\Lji$, $R_{\rm i}$, and
$\sigma_{\rm i}$ for the C+O star models with metallicity $Z=0$ and
$0.02$. We adopt progenitor stars constructed from C+O cores of 
$40M_\odot$ stars with $Z=0$ and $Z=0.02$ \citep{Umeda05} by attaching C+O 
envelopes in hydrostatic and thermal equilibrium. The envelope connects
to the core structures continuously and smoothly in the first order
differentials (e.g., \cite{Saio88}) and extends down to a density
$\rho=10^{-10}~{\rm g~cm^{-3}}$. 

The density at the O layer
is one order of magnitude lower in the model with $Z=0.02$ than in the model
with $Z=0$. The O layer can be divided into two layers according to the
abundance of C and Mg, O+Mg and O+C layers.
The boundaries between the two layers for the models with $Z=0$ and $Z=0.02$
are $3\times10^{9}~{\rm cm}$ and $8\times10^{8}~{\rm cm}$, respectively.

The parameter ranges of the outflow are
as follows; (1) We adopt $\Ljie=\Lji/(10^{52}~{\rm erg~s^{-1}})=(4,~10)$ 
because the GRB isotropic luminosity is normally-distributed in the
range of 
$L_\gamma^{\rm iso}\sim 10^{51-54}~{\rm erg~s^{-1}}$ \citep{Ghirlanda10}.
(2) $R_{\rm i}$ corresponds to the size of central engine that
is constrained by time variability of GRB prompt emission. Since an
interval of the time variability $\Delta t$ is of the order of $10^{-3}$ s,
$R_{\rm i}$ is presumed to be smaller than 
$c \Delta t \sim 10^8~{\rm cm}$. Therefore, we vary $R_{\rm i}$ for a
range of $R_{\rm i,7}=R_{\rm i}/(10^7 ~{\rm cm})=1\sim10$.
(3) $\sigma_{\rm i}$ is
determined by the mechanism to launch relativistic jets. We construct models
with $\sigma_{\rm i}=0$, which corresponds to the standard fireball model, and
with a range of $\sigma_{\rm  i}=0.1\sim 10$, which corresponds to magnetized
jet models. We note that the strength of magnetic field realizing adopted
$\sigma_{\rm i}$ is $\sim10^{12-15}$ Gauss at $R_{\rm i}$.
(4) For the other parameters, we fix $\Gamma_{\rm max}=100$ referring to a
requirement to avoid compactness problem \citep{Piran04}, and
expediently set $\epsilon \Omega=10^3~{\rm s^{-1}}$ assuming
$\epsilon=0.1$ and $\Omega=10^4~{\rm s^{-1}}$. We note that 
$\epsilon \Omega$ is rather uncertain. Initial Lorentz factor of
the outflow is fixed to be $\Gamma_{\rm i}=1.23$, which corresponds to the
sound velocity of ultra-relativistic fluids $c_{\rm s}=1/\sqrt{3}$. 
 We name the model with $(L_{\rm j,52}^{\rm iso},~Z)$=(4, 0) Model A,
  $(L_{\rm j,52}^{\rm iso},~Z)$=(10, 0) Model B, $(L_{\rm j,52}^{\rm
    iso},~Z)$=(4, 0.02) Model C, and $(L_{\rm j,52}^{\rm iso},~Z)$=(10, 0.02)
  Model D. The model parameters are summarized in Table \ref{tab:model}.

In the two-dimensional relativistic hydrodynamics calculation, we set
parameters of the relativistic jets corresponding to the parameters of
GRB outflow. The
relativistic jets are injected from the inner boundary at an enclosed
mass $1.4M_\odot$ corresponding to a radius $R_{\rm in}=900$~km. The
initial half angle of jet is set to $15^\circ$
\citep{Zhang03} and thus energy deposition rates of jets $\Lj$
is set to be $\Lje=\Lj/10^{51}~{\rm erg~s^{-1}}=1.5$ and $3.4$, which give
$\Ljie=4$ and $10$, respectively. We employ the Lorentz factor of the jet
at the inner boundary as $\Gamma=2.5$ by reference to \S~\ref{sec:outflow}
and the energy density of the jet as to accelerate the jet to $\Gamma=100$.

\section{RESULTS}\label{sec:results}

\subsection{Initial composition}
\begin{figure}
\begin{center}
\FigureFile(75mm,180mm){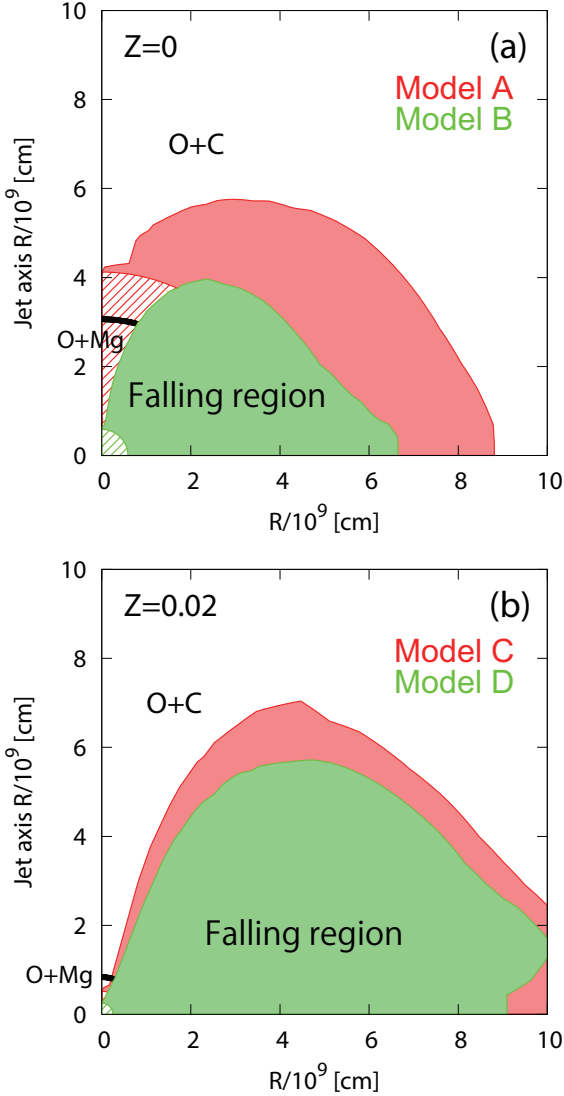}
\end{center}
\caption{Initial locations of the falling mass elements (filled and
 shaded regions), for (a) $Z=0$ models  (Model A and B) and (b) $Z=0.02$
 models  (Model C and D). The shaded and filled regions represent the mass
 elements fell 
 before and after the initiation of the jet injection, respectively. 
 We assume that the outflow consists of the materials initially at the filled
 regions. 
 The color of the regions represent the models with  $\Ljie=4$
 (green) and  $\Ljie=10$ (red) and the background circles represent the boundaries
between an inner O+Mg layer with 
$X({\rm ^{24}Mg})>X({\rm ^{12}C})$ and an outer O+C layer with
$X({\rm ^{24}Mg})\leq X({\rm ^{12}C})$.}
\label{fig1}
\end{figure}

Figures~\ref{fig1}a and \ref{fig1}b show the falling regions of the
models. The matter at the intersection between the circumference of the
falling region and the jet axis is located at the inner boundary when
the jet injection initiates.  In the models with $Z=0$  (Model A and B),
the jet injection 
initiates at early time for the model with  $\Ljie=10$ (Model B), while it
has to wait until the large portion of the material in the O layer falls in 
the model with  $\Ljie=4$ (Model A) because the ram
pressure of the jets cannot overcome that of the falling matter in 
the O layer (see \cite{Maeda09}). As a result, the falling region of the 
model with  $\Ljie=4$ is more extended than that of the model
  with  $\Ljie=10$.
On the other hand, in the models with $Z=0.02$  (Model C and D), due to
the low-density O layer the jet injections in both of the models with
 $\Ljie=4$ and $10$ initiate at similar epochs and thus their
  falling regions are similar. The initial composition of the GRB outflow is
  derived from the integration of matter at the filled region because the
  matter at the shaded region is likely to be accreted to the central remnant
  before the initiation of the jet injection.

\begin{figure}
\begin{center}
\FigureFile(80mm,80mm){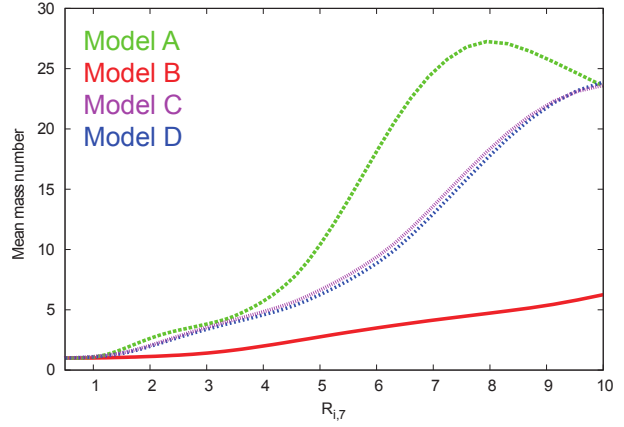}
\end{center}
\caption{Mean mass number of initial composition of the outflow.} 
\label{fig2}
\end{figure}

The composition of a mass element falling to $r=R_{\rm i}$ depends on
the maximum temperature that is higher for smaller $R_{\rm i}$ and for
the matter locating initially at the inner layers. 
The mean mass numbers $\langle A\rangle$ of the initial composition of the
outflow are shown as a function of $R_{\rm i}$ in Figure~\ref{fig2}.  The
Model B
has small $\langle A\rangle=1-6$ for $R_{\rm i,7}=1-10$ because the
jets consist of the matter initially locating in the inner layer, while 
  the Model A can have $\langle A\rangle\geq15$ for $R_{\rm i,7}\geq5.5$
because the heavy nuclei can survive in the outer matter that falls to
the outer $R_{\rm i}$. The most abundant nucleus in the models with 
$\langle A\rangle\geq15$ is $^{28}$Si due to the O burning during the fallback. 
The turnover at 
$R_{\rm i,7}\sim8$ stems from the fact that the maximum temperature of the
mass elements in the outer layer is not high enough to ignite $^{16}$O.

On the other hand, the jet injection in the models with $Z=0.02$  (Model C
  and D) are initiated at early time and the outflow contains the matter initially
locating in the O+Mg layer. The falling region are similar
in both models and thus resultant $\langle A\rangle$ of the models are
similar. In the models with $Z=0.02$, the presupernova temperature of
the infalling matter is lower than in the models with $Z=0$ due to the
low-density O layer. Therefore, $\langle A\rangle\geq15$ is
realized at $R_{\rm i}\geq7.3$ and the most
abundant nucleus is $^{28}$Si in these models.

\subsection{Outflow dynamics}
\label{sec:outflow}

\begin{figure}
\begin{center}
\FigureFile(80mm,80mm){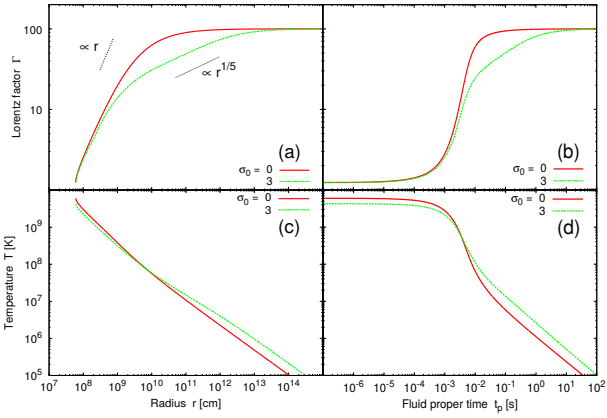}
\end{center}
\caption{Hydrodynamical properties of the models with $(\Ljie, R_{\rm
  i,7},\sigma_{\rm i})=(4,6,0)$ and $(4,6,3)$: (a) radial evolution of Lorentz
  factor, (b) time evolution of Lorentz factor, (c) radial evolution of the
  temperature, and (d) time evolution of the temperature.}
\label{fig3}
\end{figure}

Hydrodynamical properties of the outflow for the models with $(\Ljie, R_{\rm
  i,7},\sigma_{\rm i})=(4,6,0)$ and $(4,6,3)$ are shown in Figures
\ref{fig3}(a)-\ref{fig3}(d).

Figures \ref{fig3}(a) and (b) show evolution of the Lorentz factor as
functions of $r$ and fluid proper time $t_{\rm p}$ which is calculated by an
integration $t_{\rm p}=\int_{\rm R_i}^{r}1/(v\Gamma) dr$, respectively. In the
model with $\sigma_{\rm i}=0$, the Lorentz factor evolves linearly with
radius until it reaches $\Gamma_{\rm max}$ as expected from the standard
fireball model. The acceleration takes place at $t_{\rm p}\sim 10^{-3} -
10^{-2}~{\rm s}$. On the other hand, in the models with $\sigma_{\rm i}=3$,
the outflow is accelerated initially by thermal pressure like the fireball
model and later by magnetic field with the time scale of
$\tau_{\rm dis}$ . The Lorentz factor evolves more slowly than the model with
$\sigma_{\rm i}=0$. 

Figures \ref{fig3} (c) and (d) show evolution of the temperature as
functions of $r$ and $t_{\rm p}$, respectively. In the both models with
$\sigma_{\rm i}=0$ and 3, the temperature decreases exponentially with $t_{\rm
  p}$, during the phase in which the Lorentz factor evolves
linearly with radius. This stems from the relations, $\Gamma\propto r$,
$T\propto r^{-1}$, and $dt_{\rm p}=dt/\Gamma\sim dr/c\Gamma$. While the
temperature in the model with $\sigma_{\rm i}=3$ is lower than that in the
model with $\sigma_{\rm i}=0$ before the acceleration, it decreases more
slowly and becomes higher at $r>10^{10}$ cm and $t_{\rm p}> 10^{-2}~{\rm s}$ 
than in the model with $\sigma_{\rm i}=0$. This is because the acceleration is
slow and the magnetic field energy is converted not only to the kinetic energy
but also to the thermal energy in the model with $\sigma_{\rm i}=3$.

\subsection{Nuclear composition}

\begin{figure}
\begin{center}
\FigureFile(80mm,160mm){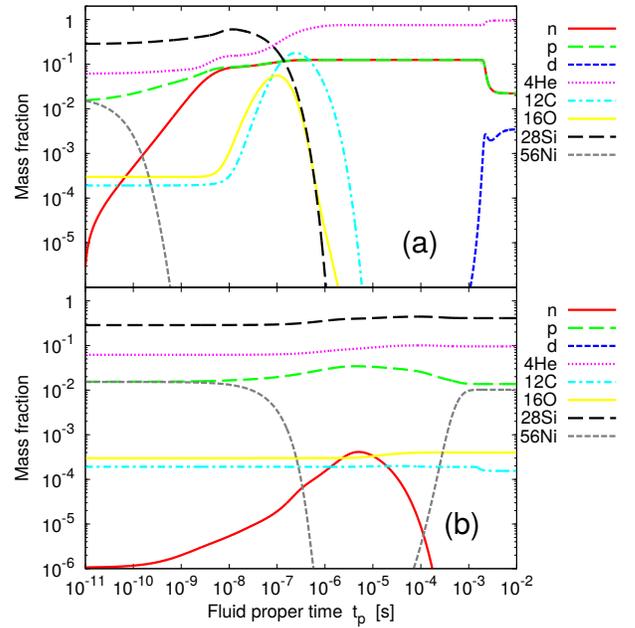}
\end{center}
\caption{Time evolution of nuclear composition for the models with
  (a)$(\Ljie,Z,R_{\rm i,7},\sigma_{\rm i})=(4,0,6,0)$ and (b)$(\Ljie,Z,R_{\rm
    i,7},\sigma_{\rm i})=(4,0,6,3)$.}
\label{fig4}
\end{figure}

Figures \ref{fig4} (a) and (b) show time evolution of nuclear
composition of the outflow for models with $(\Ljie,Z,R_{\rm i,7},\sigma_{\rm
  i})=(4,0,6,0)$ and $(4,0,6,3)$, respectively. Nuclear reaction ceases at
$t_{\rm p}\sim10^{-2}$ sec in both models (a) and (b) because the temperature
after the epoch falls below $\sim 10^8$ K (Figure \ref{fig3}d). The epoch
corresponds to $r\simeq10^{10}$ cm for the model with $(\Ljie,Z,R_{\rm
  i,7},\sigma_{\rm i})=(4,0,6,0)$ and $r\simeq4\times10^{9}$ cm for the model
with $(\Ljie,Z,R_{\rm i,7},\sigma_{\rm i})=(4,0,6,3)$. 

In the model with $ (\Ljie,Z,R_{\rm i,7},\sigma_{\rm i})=(4,0,6,0)$, which
corresponds to the standard fireball model, heavy and intermediate mass nuclei
are almost dissociated until $t_{\rm p}\sim10^{-5}$ s and $^4$He and d are
synthesized at $t_{\rm p}> 10^{-3}$ s as the outflow cools. Only light nuclei
and free nucleons, $^4$He, d, p, and n, remain in the outflow after
the acceleration (Figure \ref{fig4}a).
This demonstrates that the standard fireball model with $(\Ljie,Z,R_{\rm
i, 7},\sigma_{\rm i})=(4,0,6,0)$ destroys heavy nuclei even if they are 
initially contained in the outflow.

On the other hand, in the model with $(\Ljie,Z,R_{\rm i,7},\sigma_{\rm i})=(4,0,6,3)$,
whereas some nuclei such as $^{56}$Ni are dissociated at $t_{\rm
  p}<10^{-6}$ s and re-synthesized at $t_{\rm p}>10^{-3}$ s,
other nuclei such as $^{28}$Si survive without dissociation and 
remain abundant after acceleration (Figure \ref{fig4}b).
This illustrates that the metal nuclei can survive in the magnetized jet. 

\begin{figure*}
\begin{center}
\FigureFile(160mm,160mm){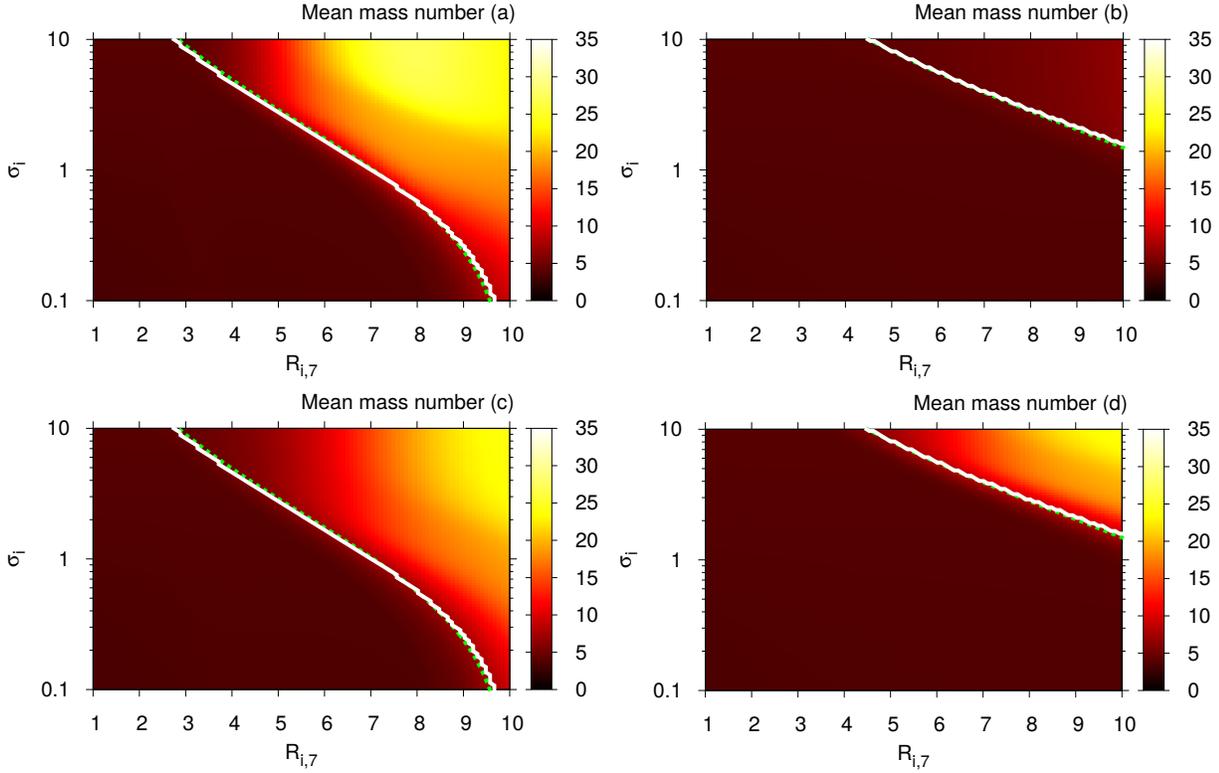}
\end{center}
\caption{Mean mass number of the final nuclear composition in the outflow as
  functions of $R_{\rm i}$ and $\sigma_{\rm i}$  for (a) model A, (b) model B,
  (c) model C, and (d) model D.
  The model parameters are summarized in Table \ref{tab:model}. 
  The white solid lines and green dotted lines represent the criterion for the
  QSE establishment and the contour of $T_{\rm i,9}=4.7$, respectively.} 
\label{fig5}
\end{figure*}

Figures \ref{fig5}(a)-\ref{fig5}(d) show mean mass number of final nuclear composition as
functions of $R_{\rm i}$ and $\sigma_{\rm i}$ for the model A, B, C, and D
(see Table \ref{tab:model}).

When a condition
\begin{equation}\label{eq:QSE}
 T_9> 3.6\left(\frac{0.04}{t_{\rm p}}\right)^{\frac{1}{33.3}}
\end{equation}
is satisfied, quasi-statistical equilibrium (QSE) is attained
(e.g., \cite{WAC73}). The white solid line in Figures
\ref{fig5}(a)-\ref{fig5}(d) represents the 
criterion to establish QSE. Therefore, in the models with lower $\sigma_{\rm
  i}$ and  smaller $R_{\rm i}$ than the white solid line, the composition can 
be described by QSE and the metal nuclei are almost destroyed due to the high
entropy. This criterion is well fitted by a contour of initial temperature
$T_{\rm   i,9}=T_{\rm   i}/10^9{\rm K}=4.7$ (green dotted line in Figure 
\ref{fig5}). On the other hand, as long as QSE is not attained, the models
with larger $R_{\rm i}$ yield final compositions with heavier mean mass number
except for the model B that initially does not have heavy nuclei
(Figure \ref{fig2}).

\begin{table*}
\caption{The 10 most abundant nuclei and their mass fractions for the models A, B, C, and D with 
($R_{\rm i}$, $\sigma_{\rm i}$)=(10, 10).}
\begin{center}
\begin{tabular}{c|rc|rc|rc|rc}
  \hline
  \hline
  & \multicolumn{2}{c}{Model A} &\multicolumn{2}{|c}{Model B}
  & \multicolumn{2}{|c}{Model C} &\multicolumn{2}{|c}{Model D}\\
  \hline
  1&
  $^{28}$Si $\cdots$ & 3.7182E-01&$^{4 }$He $\cdots$ & 5.1509E-01&
  $^{28}$Si $\cdots$ & 3.9266E-01&$^{28}$Si $\cdots$ & 3.8889E-01\\
  2&
  $^{16}$O  $\cdots$ & 2.9626E-01&$^{58}$Ni $\cdots$ & 9.0864E-02&
  $^{16}$O  $\cdots$ & 2.0150E-01&$^{32}$S  $\cdots$ & 1.7979E-01\\
  3&
  $^{32}$S  $\cdots$ & 2.0439E-01&$^{32}$S  $\cdots$ & 7.2190E-02&
  $^{32}$S  $\cdots$ & 1.7598E-01&$^{16}$O  $\cdots$ & 1.6323E-01\\
  4&
  $^{36}$Ar $\cdots$ & 5.3067E-02&$^{28}$Si $\cdots$ & 5.8950E-02&
  $^{54}$Fe $\cdots$ & 6.9657E-02&$^{54}$Fe $\cdots$ & 8.7961E-02\\
  5&
  $^{40}$Ca $\cdots$ & 4.1669E-02&$^{54}$Fe $\cdots$ & 5.7403E-02&
  $^{36}$Ar $\cdots$ & 3.9364E-02&$^{36}$Ar $\cdots$ & 4.0427E-02\\
  6&
  $^{54}$Fe $\cdots$ & 1.0189E-02&$^{55}$Co $\cdots$ & 3.3212E-02&
  $^{40}$Ca $\cdots$ & 2.2114E-02&$^{40}$Ca $\cdots$ & 2.4969E-02\\
  7&
  $^{24}$Mg $\cdots$ & 7.1779E-03&$^{36}$Ar $\cdots$ & 2.4232E-02&
  $^{55}$Co $\cdots$ & 1.6240E-02&$^{55}$Co $\cdots$ & 1.9738E-02\\
  8&
  $^{56}$Ni $\cdots$ & 2.8607E-03&$^{57}$Ni $\cdots$ & 1.9790E-02&
  $^{24}$Mg $\cdots$ & 1.0641E-02&$^{56}$Ni $\cdots$ & 1.4527E-02\\
  9&
  $^{55}$Co $\cdots$ & 2.1998E-03&$^{16}$O  $\cdots$ & 1.8796E-02&
  $^{58}$Ni $\cdots$ & 8.9718E-03&$^{58}$Ni $\cdots$ & 1.3446E-02\\
  10&
  $^{58}$Ni $\cdots$ & 1.2994E-03&$^{56}$Ni $\cdots$ & 1.4453E-02&
  $^{56}$Ni $\cdots$ & 8.8717E-03&$^{4 }$He $\cdots$ & 1.0203E-02\\
  \hline
\end{tabular}
\label{tab:frac}
\end{center}
\end{table*}

Table \ref{tab:frac} shows the 10 most abundant nuclei and their mass fractions for 
the model A, B, C, and D with ($R_{\rm i}$, $\sigma_{\rm i}$)=(10, 10).
In the model A, C, and D, $^{28}$Si is most abundant and its mass fraction reaches
$\sim40\%$. The mass fractions of $^{16}$O and $^{32}$S are also high in the model 
A, C, and D. In particular, the mass fraction of $^{16}$O reaches $\sim30\%$ 
in the models A.
On the other hand, in the model B, the most abundant nucleus is $^4$He 
the mass fraction of which is about $\sim 52\%$ and the mass fractions of each metal nucleus
are less than $10\%$.

So far, we fix the isotropic jet luminosity as $\Ljie$=4 and 10 referring to
canonical GRB isotropic luminosity. However, there are GRBs
belonging to less energetic class called low-luminosity GRBs (LLGRBs)
with $L_{\rm j}^{\rm iso}\lesssim 10^{49}~{\rm erg~s^{-1}}$ (e.g., \cite{Liang07}).
From Equation (\ref{eq:ener}) and the equation of state, the isotropic luminosity
$L_{\rm j}^{\rm iso}$ of a GRB jet is related to the initial temperature
through $e_{\rm int}$ with Equation (\ref{eq:e_int}) as 
$L_{\rm j}^{\rm iso}\simeq 16\pi R_{\rm i}^2 e_{\rm int}\Gamma_{\rm i}^2v_{\rm
  i} (1+\sigma_{\rm i})/3$. 
Applying the condition for $T_{\rm i}$ to avoid the QSE establishment (i.e.,
$T_{\rm i,9}< 4.7$), the condition in terms of the isotropic luminosity is
obtained as follows,
\begin{equation}\label{eq:criterion}
  L_{\rm j}^{\rm iso}\lesssim 3.9\times 10^{50} R_{\rm i,7}^2(1+\sigma_{\rm
    i})~~{\rm erg~s^{-1}}.
\end{equation}
According to this condition, the metal nuclei can be involved in the
jets of the LLGRBs even within
the framework of standard fireball model (i.e., $\sigma_{\rm i}=0$). 

\begin{figure}
\begin{center}
\FigureFile(80mm,80mm){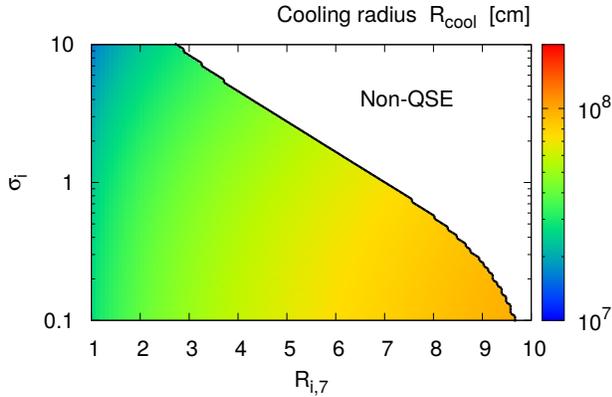}
\end{center}
\caption{The cooling radius  $R_{\rm cool}$, above which QSE is no longer
  established, as functions of $R_{\rm i}$ and $\sigma_{\rm i}$ for the model A.}
\label{fig6}
\end{figure}

Figure \ref{fig6} shows the cooling radius $R_{\rm cool}$ above which the
condition to establish QSE is no longer satisfied; $R_{\rm cool}\sim10^{7-8}$
cm depending on $R_{\rm i}$ and $\sigma_{\rm i}$. Since QSE is not attained at
$r>R_{\rm cool}$, the metal nuclei, that have not been dissociated at $r\sim
R_{\rm cool}$, can survive after the acceleration. Furthermore, if the metal
nuclei mix from the stellar mantle at $r>R_{\rm cool}$, as discussed in 
\citet{Horiuchi12}, they are involved in the outflow without serious 
dissociation.

\section{CONCLUSION}\label{sec:conclusion}
In this paper, we 
investigated nuclear composition of GRB jets assuming that the jets initially
possess metal nuclei.
We calculated fallback in a relativistic jet-induced explosion with a
two-dimensional relativistic hydrodynamics calculation and derived the initial
composition of the jet. Then, we calculated the acceleration of magnetized GRB
jets and detail nuclear reactions in the jets with the initial compositions. 

We found that the composition of the falling matter after the jet injection
depends on the radius to which the matter falls off and thus
the initial composition of the jet depends on the size of the central engine
of the jet. If the size of the central engine is larger than 
$R_{\rm i,7} \gtrsim 5$, the matter contains metal nuclei abundantly except for
the model with $Z=0$ and  $L_{\rm j,52}^{\rm iso}=10$ (model
B). Model B involves only a small fraction of metal nuclei with the mean mass
number of $\langle A \rangle \leq 6$ even for $R_{\rm i,7} \gtrsim 5$
because the jets consist of mass elements with high temperature in the
presupernova star.

We conclude that the metal nuclei can survive in the jet if QSE is not
established. The metal nuclei are dissociated mostly to $^4$He once QSE is
established. This is due to the high entropy of the jet being accelerated to
$\Gamma=100$, in which $\alpha$-rich freezeout takes place. Therefore, the
final nuclear composition of the jet is dominated by $^4$He for the
QSE-established models. The criterion for QSE establishment is well fitted 
by the contour of $T_{\rm i,9}=4.7$.  The criterion leads to the condition of
the isotropic jet luminosity $L_{\rm j}^{\rm iso}$ as $L_{\rm j}^{\rm iso}\lesssim
3.9\times 10^{50} R_{\rm i,7}^2(1+\sigma_{\rm i})~~{\rm erg~s^{-1}}$ and this
is consistent with the results of \citet{Horiuchi12} since the beaming
correction for the jet reduces $L_{\rm j}^{\rm iso}$ by the orders of $\sim 2$.

The most popular model for acceleration of GRB jet is the fireball model,
in which thermal pressure accelerates the jet from
sub-relativistic to ultra-relativistic
\citep{Goodman86,Paczynski86,Meszaros06}. In such the standard fireball 
model, $R_{\rm i,7}>5$ is required for the metal nuclei to survive in the jet
with $L_{\rm j,52}^{\rm iso} \gtrsim 1$. If the jet has a luminosity of
$L_{\rm j,52}^{\rm iso} \gtrsim 4$, the size of the central engine must be
larger than $10^8~{\rm cm}$ for the survival of metal nuclei. However, this
violates a constraint on the size of central engine from time variability
of the flux, i.e. $R_{\rm i,7}\lesssim 10$. Therefore, at least, metal nuclei
initially contained in the GRB jet with $L_{\rm j,52}^{\rm iso} \gtrsim 4$
should be destroyed and the final nuclear composition is dominated by light
nuclei and free nucleons within the framework of the standard fireball model.

On the other hand, the magnetized jet has been proposed to explain spectra of some
GRBs.
For example, \citet{ZP09} suggests that GRB 080916C involves a
magnetized jet because a thermal component, that should appear in the spectrum
if the jet energy is initially dominated by thermal energy, was not
detected. Also, \citet{Guiriec11} reports that GRB 100724B exhibits typical
non-thermal spectrum, called Band spectrum, with a significant thermal
component and suggests that a highly magnetized jet can explain the feature,
which is quite challenging for the standard fireball model.
In such a magnetized jet, the energy is initially possessed by the magnetic
field, and thus the condition for QSE is avoidable in the jet even with high
luminosities like $L_{\rm j,52}^{\rm iso} \gtrsim 10$ with satisfying 
$R_{\rm i,7}\lesssim 10$, although $R_{\rm i,7}\gtrsim 5$ is required for metal
nuclei to  initially exist in the jet in our model.

\section{DISCUSSION}\label{sec:discussion}

In this paper, we constrain the size of the central engine with the time
variability of the gamma-ray emission since it could reflect the time
variability of the central engine, for example, in the internal shock
model. However, the time variability of the gamma-ray emission may come from
other factors. For examples, the order of $\sim 1~{\rm s}$ variability can arise
from the interaction between the jet and the progenitor star \citep{MLB10},
or the time variability may be related to the emission mechanism itself,
e.g. turbulent motion at the gamma-ray emitting region (e.g., \cite{ZY11}). In
these cases, the size of the central engine is not necessarily
constrained, at least, by the time variability and the larger $R_{\rm i,7}>10$
is possible. If the size of the central engine is as large as 
$R_{\rm i,7}=50$, the criterion for the survival of the metal nuclei is 
$L_{\rm j}^{\rm iso}\lesssim10^{53}(1+\sigma_{\rm i})~~{\rm erg~s^{-1}}$,
indicating that a large fraction of GRBs can possess the metal nuclei in the
jet even with $\sigma_{\rm i}=0$.

We treated the GRB jet as a steady, radial, magnetized outflow for
simplicity. However, in reality, the jet should propagates through the
progenitor and the interaction between them is expected. Such an interaction
could affect the dynamics of the jet. For examples, it is suggested that the
initial confinement by the collapsing stellar material has an important role
on the collimation of magnetically 
dominated jets from magnetars \citep{UM07}. The cocoon, which is a shocked hot
gas surrounding the jet, also have a role to collimate the jet (e.g., 
\cite{MLB07,MA09,Bromberg11}). 
If the jet is collimated, the evolution of jet cross section $\Sigma(r)$ may be
expressed as $\Sigma(r)\propto r^{\xi}$ with $\xi<2$. In this case, the
temperature, and also the density, will decrease slowly than in the radial
flow and the criterion, Equation (\ref{eq:criterion}), possibly tighten
because the fluid will tend to keep the high temperature at the initial stage.

 \citet{Metzger11b} investigated the nucleosynthesis from free nucleons in
  the magnetically dominated jet in the context of the protomagnetar model for
  the central engine of GRBs \citep{Metzger11a}. 
  The main difference between their work and this paper is the parameter range
  of the entropy per baryon. 
  They estimated the entropy per baryon in the jet with the analytic
  expression derived in \citet{QW96} for the neutrino-driven winds.
  They suggested that, since the entropy per baryon in
  the neutrino driven wind is sufficiently low, the heavy nuclei
  beyond the iron peak, i.e., $A\gtrsim56$ where $A$ is a mass nubmer, can be
  synthesized in the jet.
  On the other hand, we do not consider such sufficiently low entropy
  environment for synthesis of metal nuclei from free nucleons and we focus
  only on the survival of the metal nuclei in this paper. 
  Future studies on the central engine of GRBs will reveal whether such low
  entropy enviornment is realized or not.

It is known that GRBs are one of the candidates for the origin of ultra high
energy cosmic rays (UHECRs) (e.g., \cite{Waxman95,Vietri95}) and LLGRBs
  are also possible candidates for the origin of UHECRs  (e.g.,
  \cite{Murase06,GZ07}).
The acceleration of metal nuclei up to ultra-high
energies could be possible in both usual high-luminosity GRBs and LLGRBs
\citep{WRM08,Murase08}.
\citet{Auger10} reported that the observed UHECRs are dominated by heavy
nuclei at high energies, i.e., $E\gtrsim 10^{19}{\rm ~eV}$ (see, however, 
\cite{HiRes10}). \citet{HT10} suggested that the result of \citet{Auger10} can
be quantitatively reproduced only if UHECR composition at the accelerating
site mainly consist of intermediate mass nuclei such as nitrogen together with
a considerable fraction of heavy nuclei such as iron. Therefore, GRBs with
magnetized jets or LLGRBs have a potential to be the origin of UHECRs in terms
of, at least, the nuclear composition because intermediate mass nuclei, 
e.g. $^{28}$Si, are very abundant in the jet.
 However, in order to explore whether the model considered in this paper can
quantitatively explain the UHECR observations,
acceleration process, escape from the source, and the propagation of the UHE
nuclei from the source to the Earth also have to be considered.
Such calculations are beyond the scope of this paper and will be studied
elsewhere.

\section*{Acknowledgments}
We thank Hajime Susa for his helpful comments. This research has been
supported in part by World Premier International Research Center Initiative,
MEXT, Japan, and by the Grant-in-Aid for Scientific Research of the JSPS
(23740157, 25$\cdot$2912). Data analyses were in part carried out on the general-purpose PC
farm at the Center for Computational Astrophysics, CfCA, of National
Astronomical Observatory of Japan.

\end{document}